# Complimentary Phylogenetic Signals for Morphological Characters and Quantitative 3D Shape Data within genus *Homo*

Peter J. Waddell[1]

pwaddell.new@gmail.com

[1]Ronin Institute, 1657 Upland Dr, Columbia, SC 29 204

Estimating the phylogeny of the genus *Homo* is entering a new phase of vastly improved data and methodology. There is increasing evidence of at least 4 to 10 competing species/lineages at any point in the last half million years, making the elucidation of the relationships of individual specimens particularly important. Recent estimates of the phylogeny of key specimens include Waddell (2013, 2014, 2015, 2016), and Mounier et al. (2016). These are made with quite different data (3D skull shapes and discrete morphological characters, respectively) and methods of analysis (unweighted least squares fitting of distances, OLS+, and reweighted maximum parsimony, respectively). Initial inspection of the trees in these articles might leave the impression of major disagreement. Here it is shown this need not be the case, and that these two types of data and analysis may be indicating a very similar tree, one that is in good agreement also with subjective current wisdom/expert opinions on particular parts of the phylogeny. The precise location of the African LH18 specimen arises as key to a better understanding of the likely form of the last common ancestors of *H. sapiens* and Neanderthals. A diverse approach seems to bring forth much more agreement of trees than otherwise perceived, and argues against being dogmatic about methods of phylogenetic analysis particularly when working with difficult problems.

This version 12/30/2016

# 1 Introduction

It is possible that until the last 100 kya perhaps 6-10 distinct ancient lineages within the genus *Homo* acted like species in undergoing quite separate evolution and adaptation (Waddell 2016). Some of these are relatively well known (*H. sapiens*, *H. neanderthalensis*), while others are known almost exclusively from their near complete sequenced genomes (the Denisovan lineage, Reich et al. 2010). Yet others have left tantalizing hints that they are not any of the above based on morphology, e.g. *Homo floresiensis (*Brown et al. 2004), *H. iwoelerueensis*-the Iwo Eleru skull (Brothwell and Shaw 1971, Havarti et al. 2010, Waddell 2014), the Boskop and Fishhoek skulls (Schwartz 2016), the Sale specimens (Schwartz and Tattersal 2003), Skull 5 (Schwartz 2016), and the Red Deer people, Curnoe et al. 2012).

Some of the most suggestive evidence for long surviving distinct lineages are biogeography and age (that is, locations in space and time). The situation observed in the last 100kya was also appears to have been the situation earlier in time. This is not only because of the assumed antiquity of some of these lineages (e.g., *Homo floresiensis*), but also the presence of forms close in space and time that do not fit easily together. For example, in the last 200-450 kya Europe has seen forms as diverse as the Sima de los Huesos specimens (about 400 kya), the Steinheim skull (~250-300kya), the Ehringsdorf skull (230kya), the Petralona skull (150-250kya), Saccopastore 1 (~100-130kya, but perhaps ~250kya) and the Ceprano skull (~400kya). Recent phylogenetic reconstructions, genetics in the case of Sima de los Huesos (Meyer et al. 2016) and morphology for Steinheim and Ehringsdorf (Mounier and Caparros 2015, Mounier et al. 2016), have associated the first three as branching on the Neanderthal lineage, although not necessarily direct ancestors of Neanderthals.

The form of Petralona is particularly distinct and if the estimated age is right, this suggests another species in South Eastern Europe in the last 250kya. This is reinforced by the recent genetic findings confirming that hints in the morphology of the Sima de los Huesos specimens , but lacking in Petralona, are indeed synapomorphies of the lineage that included Neanderthals as its most recent known forms (Meyer 2016). It is even possible that Petralona is a Denisovan lineage skull, but it might also be sister to both Neanderthals and Denisovan lineages (Waddell 2015), or an earlier lineage (it tends to locate on estimated trees mostly in the later position, Mounier and Caparros 2015, Mounier et al. 2016). Only the geographic boundaries of the species *H. neanderthalensis* are even roughly known, and over the last ~60-100 kya, they seem to have roamed widely over Europe, the Middle East and across Eurasia into at least central southern Siberia.

Mayr's view that *H. sapiens* was basically the only species that has ever existed in the genus *Homo*, and that it was essentially a ~2 million year long chronospecies, was a useful counterpoint to an earlier assumption of many lineages, and perhaps even species, even amongst living humans. The later served the socially useful role of dampening down highly racist views of who were competent humans. The timing was unfortunately after that horse had well and truly bolted. Unfortunately, this useful scientific counterpoint, was put so strongly by a socially dominant group of scientists that it became a type of dogma (Wolpoff 1999). It was only really rocked by the findings of Wilson and colleagues (Cann et al. 1987, Vigilant et al. 1991), that then new genetic data strongly suggested that the whole species *Homo sapiens,* made up of all living humans, was both very recent (less than 200kya) and that is arose in a very specific region of the world (Africa, and even then hints that it was only from a part of sub-Saharan Africa, perhaps the size of a large country, Waddell and Penny 1996). It is important to precisely define *Homo sapiens* in phylogenetic discussions, and herein it is only the last common ancestor of living humans (and all its descendants, see Waddell 2016). This definition is reinforced by the latest genetic data.

Recent genetic findings may even be suggesting that near total replacement of lineages/species of the genus *Homo* was not just a rare and very recent phenomenon in the genus, but might even have been closer to the norm. That is, lineages/species in *Homo* arose and evolved adaptations within an effective population size of a few thousands (markedly less than that of many sub-species of great apes), and if successful, spread out, competed with and replaced other lineages. So far, the evidence is that this replacement lead to low levels (typically less than 5%) of long term effective introgression (Waddell 2016). The low effective population sizes suggest origins in areas that were closer to the size of typical large countries rather than whole continents. Interbreeding in contact zones was not necessarily rare (Waddell 2016), but often gene flow was limited. Highly individually advantageous genes, such as key active components of the immune system (e.g., HLA genes) seem to have introgressed successfully, as expected. However, hybrids often did not seem to flourish as mixing up all the genetic variants of two lineages, such as *H. neanderthalensis* and *H. sapiens*, caused problems. This was apparently both short term disruption (e.g., male fertility deficits, Sankararaman et al. 2014), but also long term disruptions such that hybrids were not more competitive than both of the parental species (Fu et al. 2016). At present, this view is being confounded by the fact that Neanderthals and Denisovans seem to have evolved in the last couple of hundred thousand years at an effective population size of less than 5,000, whereas most living *H. sapiens* populations have effective population sizes over the last 100,000 years of > 10,000 (e.g., Kuhlwilm et al. 2016). This could have resulted in a build up of deleterious mutations, that made Neanderthal and Denisovan DNA somewhat toxic. However, it is harder to see how that would explain lack of mixing in earlier instances, such and Neanderthal into Denisovan, or perhaps *erectus*-like into Denisovan, since in these cases, the effective population sizes of these lineages would be expected to be more similar.

The arguments in paleoanthropology about relationships of fossils are just now moving from one of what different camps of people might believe about human evolution, to truly being one of considerable data in a time of improving phylogenetic methodology. Methodological phylogeneticists know that every base in the genome of all members of the genus *Homo* has evolved according to a specific tree. If all these trees are samples from the same species/population tree (or close to it) that makes phylogenetic reconstruction easier. If this is not the case, there are tools to discover specific instances of hybridization/introgression, but this adds a much greater level of complexity which can be very taxing with limited data. However, if the phylogeny of *Homo* is ever to be determined, then it is necessary to start to trace the history of each specimen, and only, very cautiously, explore grouping specimens for phylogenetic analysis. Fortunately, there are tools in current software to allow different types of reversible grouping of specimens, although they are not as flexible as they could be. Many programs will allow forcing a set(s) of specimens to be monophyletic, while other options will allow constraints on the sub-tree defined by a subset of specimens (Swofford 2002). However, both types of constraint cannot be be enacted at the same time. This could be particularly useful in cases of mixed information of limited reliability.

Fortunately, a wide range of different types of data are now appearing to assist with the difficult task of inferring the phylogeny of *Homo*. One of the most well known is genomic data. A recent review of what this seems to be telling us is Waddell (2016). Also contributing is the ongoing work to better understand and evaluate the age of specimens. There is however more uncertainty than usually expressed in individual articles and specimens. Millard (2008) gives a useful overview of some of the major uncertainties that still face this area. Thus, placing too much faith on the exact date of a specimen when trying to infer a phylogeny can lead to failed hypotheses.

Recent examples of relatively large and widely scored morphological matrices for individual specimens, are Mounier et al. (2016) and Zeitoun et al. (2016). Scoring as many characters as possible on individual specimens, is the correct approach; specimens can always be grouped at some later point in the analysis and their states homogenized, but the opposite is not possible. Effects of the final result with and without grouping should also be considered. There have also been encouraging results coming from phylogenetic analysis of 3D metric shape data, via Procrustes distances (Waddell 2014, 2015, 2016). There is also considerable overlap in these three data sets. The data of Mounier et al. (2016) totals 70 characters, of which about half are endocranial and the rest on the exterior of the cranium.

Another source of information on the relationships of different specimens is current perceived wisdom including expert opinion (e.g., Rightmire 1998, Schwartz 2016, Stringer 2016). For example, all these authors express the opinion that the LH18 specimen (Magori and Day. 1983) is on the lineage to living *H. sapiens* after it split from *H. neanderthalensis*. These beliefs can be a fruitful place to start to construct informative Bayesian prior probabilities (Stuart and Ord 1987). One danger is that they can be utilizing the same data (morphological features including general shape) that are expressed in the data itself. As an alternative, they can be seen as the output of a black box method of phylogenetic analysis. In either case, congruence of hypotheses is can guide the search for the most appropriate methods of analysis. What is crucial is that the data being used as input to a phylogeny estimation program is as objective and appropriately encoded as possible.

While it is very important to have comprehensive data sets for all important data sets coded and widely agreed upon, in order to obtain good estimates of the true phylogeny, it is useful to bring to bear the full range of well researched phylogenetic methods. There are comments appearing that suggest that in the original analyses of some of these emerging data sets there  misunderstanding of methodology and, in some cases, a tendency to dogma. To quote (Mounier and Caparros 2015) "... in the case of palaeogenomic studies, on genetic distance similarity algorithms calibrated to uncertain molecular clocks, both approaches failing to take homoplasies, apomorphies and plesiomorphies into account numerically in a phylogenetic context."  This is both a misunderstanding of the basics of how genomic data sets are being analyzed and it misses the point that the biggest problem with these genetic data sets is that they, so far, only cover a few of the most recent fossils (with nuclear sequences from the Sima de los Huesos site of ~400kya being a major step forward, Meyer et al 2016). That the same authors then proceed to analyze their valuable data with only an iteratively reweighted parsimony method, that has not been shown to be more robust or efficient than other applicable methods (Swofford et al. 1996, Felsenstein 2004), is unfortunate.

When dealing with data sets of limited reliability, with parts of the data that can be biased and misleading, looking at how a wide range of tree building methods and treatments of the data work can be enlightening (Kim 1993, Waddell et al. 1999). One of the useful statistics to keep in mind is that recurrence of a group of more than two OTU's can be very rare by chance alone. In the case of the newly discovered superorder of mammals, Laurasiatheria, it was the similarity of a group encountered across multiple genes, even with weak overall bootstrap support that was most indicative this was a true clade (Waddell et al. 1999). There are also some interesting results in the literature suggesting that the consilience of two of the three main classes of evolutionary tree inference, likelihood, parsimony and distance methods, can be a better indicator of the correct group than anyone (Kim 193). These results has never been fully explained, explored or exploited. It is with thoughts such as these in mind, along with the notorious difficulty of determining how fossil specimens within the genus *Homo* are related, that helps to drive this exploration of the very recently presented data set of Mounier et al. (2016).

## 2  Materials and Methods

The first data used here is a set of 70 discretely encoded morphological characters for 28 specimens (Operational Taxonomic Units or OTU's) presented in table 3 of Mounier et al. (2016). Because it was not possible to reproduce the results in that paper, a Nexus file of the data was sought and received (A. Mounier pers. comm.). Some specific differences were then identified and their origins clarified. A subset of the Procrustes distances in Waddell (2015), based on 3D morphometric measurements from19 landmarks and 78 semi-landmarks from Harvati et al.

(2011), was used for comparison.

A number of programs and scripts were used for these distance analyses. These include PAUP* (Swofford 2002) (alpha versions 148-150 released in 2016) for all parsimony and distance-based tree inference and character based bootstrapping. For least squares residual resampling methods (Waddell, Azad and Khan 2011) Perl scripts from Waddell, Tan and Ramos (2011), were used.

## 3 Results

Here, results of maximum parsimony (unordered Wagner) analyses of the discrete morphological character-based data of Mounier et al. will be contrasted with the trees of Waddell (2014, 2015, 2016). This is followed by reanalysis of a matching subset of specimens which are identical (bar one substitute specimen) for both these quite different data sets. Finally, the impact of recoding just the first character of Mounier et al. (2016) will be assessed.

### 3.1 Maximum parsimony trees from table 1 of Mounier et al. (2016)

The data set from table 3 of Mounier et al. (2016) yields the set of MP trees indicated in figure 1a below. Overall it is in good agreement with the trees of Waddell (2014, 2015, 2016) constructed with Procrustes distances. Like those distinct analyses, this data matrix is quite strong in placing Kabwe outside of an association that is basically from the Neanderthal/*H. sapiens* split forwards in time. However, it reverses the order of branching of the Kabwe and Petralona specimens. This runs against the possibility that the Petralona specimen (~150-250 kya from Greece) may be part of the pre-Neanderthal clade (with respect to *H. sapiens*), which is one interpretation of the results in Waddell (2015).

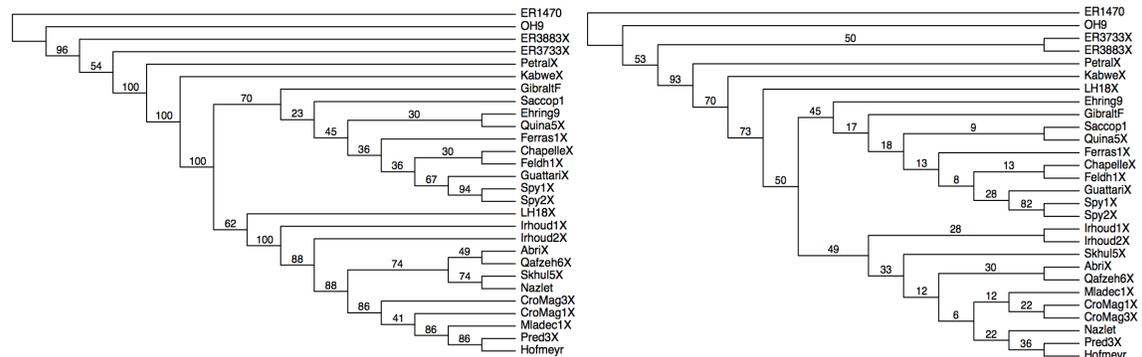

Figure 1. (a, left) The consensus tree of 176 maximum parsimony trees of length 263 for the data in table 3. (b, right) Result of 1000 bootstrap replicates of the data in table 3 of Mounier et al. (2016) analyzed with maximum parsimony, all states unordered. The tree shown is majority rule plus compatible groups. The "X" indicates specimens matching those in the trees shown in Waddell (2015, 2016).

In Mounier et al. (2016) LH18, a skull widely regarded as pre-*sapiens,* appears as pre-Neanderthal + *H. sapiens* in their preferred evolutionary tree (figure 1 of Mounier et al. 2016) obtained via iterative reweighting of the data. However, in figure 1a herein LH18 is associated as a pre-*sapiens* lineage in the majority of the unweighted maximum parsimony trees. Relative to the analyses of Waddell (2015, 2016), the exact placements of the pre-*sapiens* lineages in both figure 1a (below) and figure 1 of Mounier et al. (2016) show differences that contradict expectations. In Waddell (2015), the branching order tends to be Irhoud/LH18/Qafzeh 6/Skull 5/*H. sapiens* (with the latter here represented by Hofmeyr, Pred3, CroMag, Mladec1, Abri, and perhaps Nazlet, based on overall shape and having an inverted T-shaped chin, Schwartz 2016). Having Skull5 deepest in the pre-*sapiens* lineage in figure 1 of Mounier et al. (2016) is therefore quite different, and in a location that contrasts with that expected in Rightmire 2016, Springer 2016, and Swartz (2016). Qafzef 6 is also diagnosed as non-*H. sapiens* in Schwartz (2016), and it is recovered as a likely hybrid in Waddell (2014), but with the data of Mounier et al. (2016) it appears within the collection of *H. sapiens* skulls. However, the trees in figure 1 and in Mounier et al. (2016) do place the Neanderthals in a monophyletic group, and it seems probable that the Ehringsdorf skull is indeed sister, given its age and location deep within Europe. This is almost certainly a more accurate reconstruction than the unconstrained tree of Waddell et al. (2015), where Neanderthals are a para or even poly phyletic group. Deep in the tree of figure 1 of Mounier et al. (2016), the placement of OH9 closer to the root than 3733 and 3883 is unexpected, given the age, biogeography and endocranial volume of this specimen.

It is also interesting that the tree in figure 1a, places Irhoud 1 deeper than Irhoud 2, in agreement with the order in Waddell (2015, 2016), but opposite to that in figure 1 of Mounier et al. (2016).

The combination of bootstrapping and MP is interesting and seems to often yield results that are more reasonable than the MP tree(s) alone. Bootstrapping is, itself, described as a Bayesian method (Waddell et al. 2004), while MP itself is a maximum likelihood method. Thus, the combination might have desirable elements of a Bayesian method in better integrating support near boundaries in the parameter space of edge lengths. In this case, the bootstrapped MP tree does three appealing things; these are placing the Irhoud skulls into a monophyletic group, likewise for the CroMagnon skulls, and placing Skhul5 sister to the *H. sapiens* specimens (all consistent with expectations in Schwartz and Tattersal 2003). However, it does place LH18 deeper in the tree.

**3.2 An error in table 3 and results with the corrected data**

There is a disagreement between the data in table 3 of Mounier et al. (2016) and the data run in the actual paper. A discrepancy was found by the author when trying to replicate the analyses in that paper and in response A. Mounier sent the data as a Nexus file. Difference were noted to be due to the character states for characters 36 and 39 changing from 2 to 1 for specimens KNM-ER 1470, KNM-ER 3730, and KNM-ER 3883 (and in doing so abolishing character state 2 for these characters in any specimen). Also, character 41 changes state from 1 to 0 in 1470 and 3733, and from 2 to 1 in 3883 (again abolishing character state 2 for this character). This recoding was made following inspection of the original specimens (not casts) of these fossils (A. Mounier, pers comm.). Both forms of the data are analyzed here, but they result in relatively small differences. Results of the revised data are seen in figure 2, and again highlight the labile nature of the location of LH18 with this data. With this revised data, the RC reweighted Maximum parsimony tree obtained, although very similar to that in figure 1 of Mounier et al. (2016), does not match perfectly.

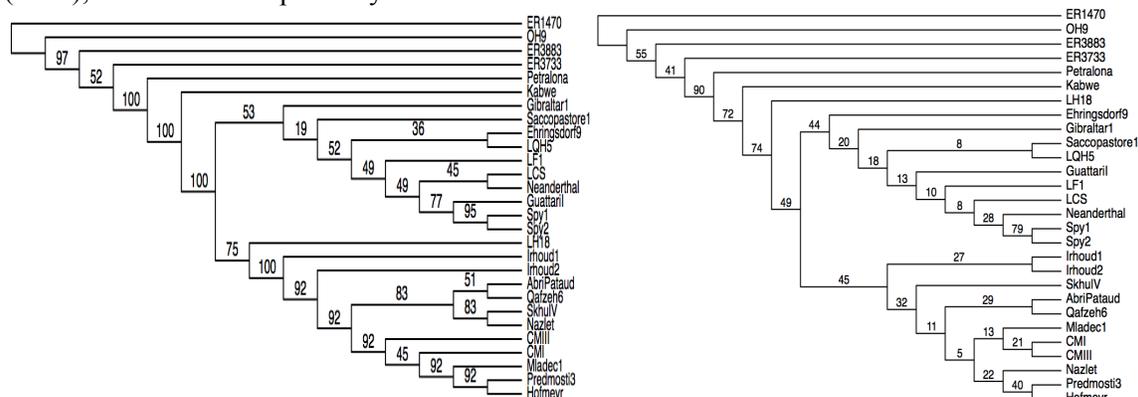

Figure 2. (a) The consensus tree of 283 Maximum parsimony trees of length 259 for all characters in the data set of Mounier (2016) as communicated by the author, found with TBR and 100,000 random addition starting trees. (b) The majority rule tree from 1000 bootstrap replicates.

It is also interesting to assess this data with OLS+ (Figure 3a) as this is a frequently used method in the phylogenetic analyses of shape data (Waddell 2014, 2015, 2016). The g%SD statistic reported by PAUP measures the degree of non-additivity of the distances on a tree (Waddell et al. 2011). Here it is 14.46% (fitted parameters, $k = 54$), which is markedly higher than seen with Procrustes distances, which were ~4 to 8% (Waddell 2014, 2015, 2016). This tree is very similar to the MP trees, keeping desirable features such as the Irhoud specimens together, but burying Qafzeh 6 more deeply within *H. sapiens* than previously. LH 18 again branches before the Neanderthal / *H. sapiens* split. However, in both figure 3a and 3b Skhul5 is in the general position expected.

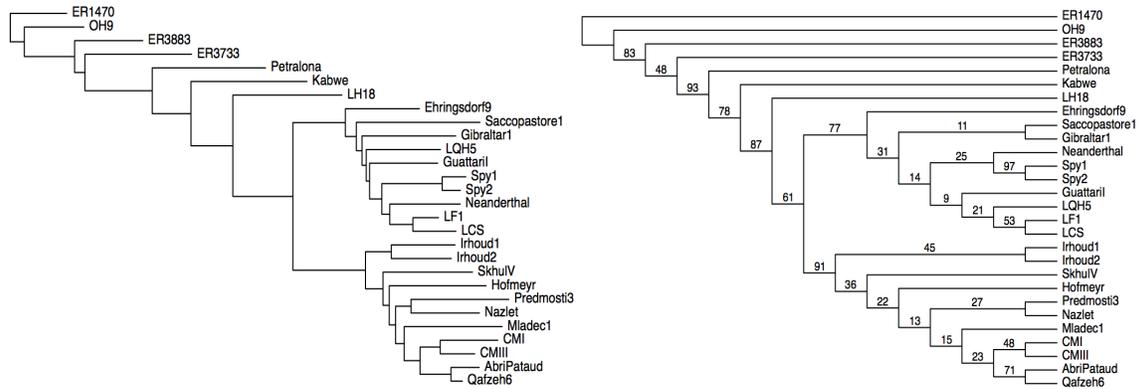

Figure 3. (a) The OLS+ tree with g%SD = 14.46 ($k = 54$) on the data of Mounier et al. (2016) as sent. (b) The majority rule consensus tree with compatible groups of character bootstrap resamples analyzed with OLS+.

**3.3 A direct comparison to OLS+ Procrustes-distance derived trees**

It is possible to more closely compare the trees produced by discretely coded characters, with those produced from Procrustes distances ($q = 0.5$ in Waddell et al. 2014) from 3D landmarks and semi-landmarks. To do this, the intersection of specimens analyzed in Waddell (2015) and Mounier (2016) were used. Since the Hofmeyr skull in the latter paper is generally regarded as modern (Grine et al. 2007), and since its location combined with genetic data suggests it is likely part of the Khoisan group, a male Khoisan skull 44 from the former paper was used as its counterpart for these comparisons. Figure 4a shows the maximum parsimony bootstrap results for this subset of taxa. The tree is similar to that seen earlier in figure 1a. Here, LH18 is more securely placed outside of the Neanderthal/*H. sapiens* split, the CroMag specimens no longer form a group, and the relationships within Neanderthals have changed. Applying OLS+ with bootstrapping of the characters to the same data results in a very similar tree except that Qafzeh 6 is buried deeper within the *H. sapiens* group (figure 4b).

Figure 5a shows the Procrustes-distance derived OLS+ tree of 21 matching taxa (plus a Khoisan male skull added in in place of the Hofmeyr skull). The g%SD is 7.38 ($k = 41$). This tree is basically the same as trimming the extra taxa off the larger unconstrained OLS+ tree in Waddell (2015). Thus, with regards to including and excluding specimens, the data in Waddell (2015) combined with OLS+ seems stable, something seen previously in comparing the subset tree of Waddell (2014) with the tree of the superset in Waddell (2015). Like those earlier analyses, the major issue with the OLS+ Procrustes distance tree seems to be that Neanderthals do not form a monophyletic group, rather they are a paraphyletic group with Petralona and SH5 interspersed (ignoring the placement of Spy 2, for example). Spy 2 is seen in Waddell (2015) as being surprisingly strongly drawn towards *H. sapiens*, and here it moves closer to *H. sapiens* than Irhoud 1. On the positive side, LH18 remains firmly placed closer to *H. sapiens* than any Neanderthal, and both Skhul5 and Qafzeh6 are firmly outside of *H. sapiens*. The Khoisan male occupies a very similar position to Hofmeyr in the matching character data (figure 4a).

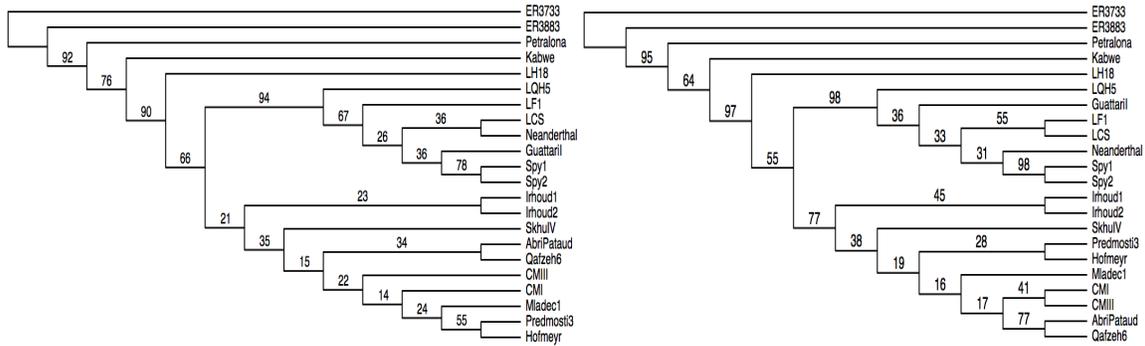

Figure 4 (a) The MP tree for the data of Mounier et al. (2016) as sent (length 203, CI 0.3655 RI 0.6429). (b) Consensus tree of OLS+ trees produced by bootstrapping character data. That is, converting characters of Mounier et al. (2016) to distances then analyzing these with OLS+. The g%SD here was 14.02 ($k = 41$).

Finally, taking the Hamming or mismatch distances produced by the data of Mounier et al. (2016) as sent, and applying residual resampling with OLS+, results in the tree of figure 5b. The support values from residual resampling in this example are well correlated with those obtained by the bootstrap of the characters, but they are generally higher. This last tree can be compared to that of figure 4b, except that residual resampling was used to assess robusticity rather than the bootstrap of characters. Here the bootstrap is introducing more uncertainty than residual resampling. However, the bootstrap is known to introduce more uncertainty than X + Yi resampling for approximating Bayesian posteriors (Waddell et al. 2004).

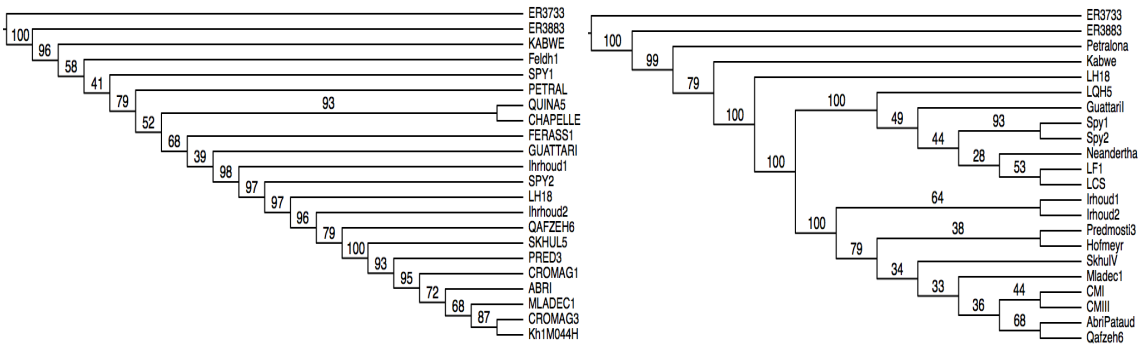

Figure 5. (a) The residual resampling OLS+ tree for matching specimens, with Hofmeyr, replaced with a Khoisan male skull (g%SD = 7.38, $k = 41$). A small ~2% upward adjustment to the resampling variance was made to compensate for the reduction in mean sum of squares in the simulated results due to tree search. (b) Character based-distances for the data of Mounier et al. (2016) as sent, with residual resampling (note, LC is Chapelle and is LF1 the La Ferrassie skull).

**3.4 Recoding just the first character**

The first character within the data set of table 3 of Mounier et al. (2016) is endocranial volume, with below 1200cc being considered underived and above 1200 being derived. The logic for choosing this particular cutoff or coding of a continuous character is not apparent. It is also problematic that this character then has a perfect fit on the MP tree (figure 1, Mounier et al. 2016). To look at the effect of this coding, which looses information, this character was recoded as stepwise binary, with endocranial volume rounded to the nearest 100cc, that is 800, 900, …, 1700 (endocranial volume data from Schoenemann 2013). To compensate for the fact this character now appears in 10 pieces, each of these is given weight 0.1 and all other characters are given weight 1. The results are shown in figure 6a. Note that the MP trees all now show

Neanderthals as a paraphyletic group, and LH18 is seen as a pre-*sapiens* lineage. The bootstrap tree is shown in figure 6b and, as seen elsewhere, here too it has produced an apparently more reasonable result with Neanderthals and Ehringsdorf reforming a monophyletic group and so to for the CroMagnon skulls.

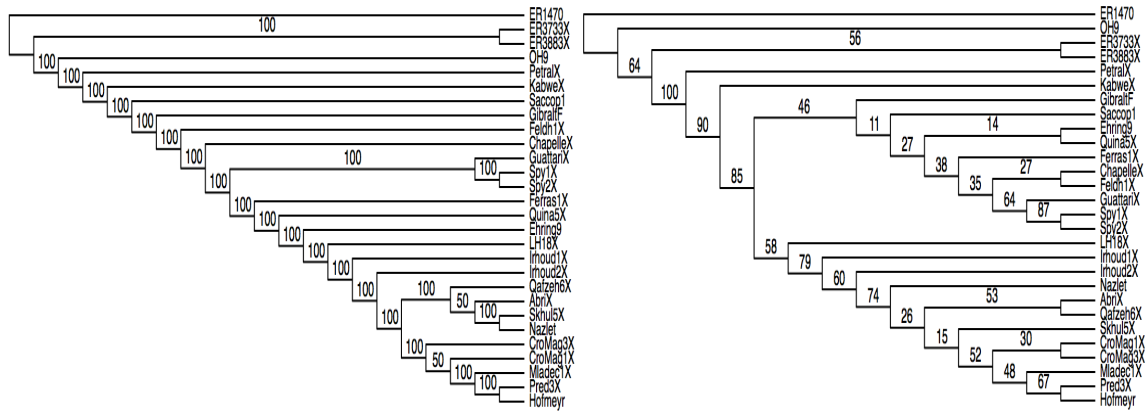

Figure 6. (a) Consensus of the shortest trees (264.5) from the data set of Mounier et al. (2016) with the first continuous character recoded in finer form. (b) The bootstrap consensus tree of this dataset.

## 4 Discussion

One useful result of this study is to show that both the discretely coded character data set of Mounier et al (2016) and trees from Procrustes derived distances based on the shape of the skull cap (Waddell 2014, 2015, 2016) are in good agreement Indeed they seem to compliment each other, and may even have similar amounts of resolving power, albeit better in different parts of the tree. For example, the discretely encoded data is capable of producing the Neanderthals as a monophyletic group, in agreement with general expectations (e.g. Schwartz and Tattersal 2002), including morphology, age and biogeography, while the shape data produces results much better in agreement with expectations for the lineage leading to *H. sapiens*.

Another useful result is showing that there can be compatibility between this data set, the Procrustes skull shape distances analyzed in Waddell (2015), and traditional morphological expectations on where LH18 should fall in the tree, That is, as an early branching member of the pre-*sapiens* lineage (Rightmire 1998, Schwartz 2016, Stringer 2016). If this is accepted, then the diagnosis of uniquely derived features (synapomorphies) for the last common ancestor of the Neanderthal and *H. sapiens* lineages needs to be reconsidered. The node marking all these specimens, plus LH18 does not appear to have any derived features that do not undergo reversions in this data. The inclusion of LH18 in the diagnosis of the features of the last common ancestors of Neanderthals and *H. sapiens* also makes this ancestor look rather more like Kabwe or Petralona specimens in terms of features. Given also the conflict between the trees in Waddell (2014, 2015, 2016) and Mounier et al. (2016) as to precise location of this specimen, it seems possible that the exact from of the ancestor of *H. sapiens* and Neanderthals will be hard to diagnose morpologically.

The results presented here argue against relying too much upon iteratively reweigthed parsimony. There are, unfortunately, no robust theoretical or simulation results to suggest it is a superior to standard Wagner Maximum Parsimony (Swofford et al. 1996, Felsenstein 2004). Indeed, it is easy to show that it will be even more strongly misleading in cases where parsimony is known to reconstruct the wrong tree (Felsenstein 1981, Hendy and Penny 1990). It is also biased towards giving higher bootstrap support to potentially misleading features of the tree if characters are not bootstrapped first and the whole analysis repeated (rather than just sampling with replacement the already estimated weights). The first aim of paleoanthropology should be to reconstruct the phylogeny and only when there is some confidence that parts of it are indeed correct, to consider which characters might best diagnose membership of specific clades or character combinations at internal nodes.

An interesting case may be building that the Petralona specimen may branch particularly close to the pre-Neanderthal lineage. The argument goes as follows. The general shape of the skull cap of Petralona in a phylogenetic analysis drops out with Neanderthals and SH5 from Sima de los Huesos in a phylogenetic analysis (Waddell 2015). SH5 has recently been shown, using nuclear ancient DNA, to most probably an early member of the Neanderthal lineage (Meyer et al. 2016). However, Petralona does not show such a clear morphological affinity to Neanderthals (e.g. signs of a suprainiac fossa). Also, if the argument of Waddell (2015, 2016) is accepted that recent results of the long term population size leading to *H. sapiens* do not seem to have experienced an early double out of Africa type population size squeeze, then that makes it look likely that many of the competitive lineages of *Homo* were arising in Africa and populating/repopulating Eurasia. Given its age and geographic location, it cannot be yet be excluded that Petralona is on the Neanderthal side of the split from modern humans. Given that we have so few clues (a tooth) as too the morphological form of the Denisova lineage, it cannot yet be excluded that Petralona might even be an early branch of that lineage. Its age (150-250 kya, Grun 1996) and biogeographic location do not argue against that possibility. If it is another member of this pre-Neanderthal lineage (as is the Denisovan) that would be a parsimonious result requiring one less migration out of Africa. If not, then there were probably multiple migrations from Africa to Eurasia in the period predating the Neanderthal modern split, ~800-500 kya.

Further exploratory analyses will be useful before looking at a larger combined analysis of all the data, perhaps in a Bayesian framework or in addition, as a weighted combination of distance matrices (e.g., Waddell 2013). One of these is the resolving power and congruence between the way Zeitoun et al. (2016) encode metric ratios (and hence the relationships between landmarks) versus the use of Procrustes, or other 3D shape-based distances, of the cranium. Also needed are far wider ranging discussions and evaluations of which characters and character encodings seem most reasonable. The results of section 3.4 clearly show that even in parts of the tree that seem strongly supported (the Neanderthal + Ehringsdorf clade), re-encoding a single character resulted in that group becoming highly paraphyletic. Another challenge will be to find the most reasonable basis on which to weight characters. This involves estimating the "variance" of such characters and their covariances or correlations. At present there are sufficient samples to do that for a number of old lineages within living humans and, perhaps also, Neanderthals.

## Acknowledgements

This work was partly supported by NIH grant 5R01LM008626 to PJW. Thanks to Kayla Miller for comments.## References

Brothwell, D., and T. Shaw. (1971). A Late Upper Pleistocene Proto-West African Negro from Nigeria. Man, (new series), 6: 221-227.

Brown, P., Sutikna, T., Morwood, M.J., Soejono, R.P., Jatmiko, S.T., Sap-tomo, E.W., and Due, R.A. (2004). A new small-bodied hominin from theLate Pleistocene of Flores, Indonesia. Nature 431, 1055–1061.

Cann, R. L., M. Stoneking, and A.C. Wilson. (1987). "Mitochondrial DNA and human evolution." Nature, 325: 31-6.

Curnoe, D., J. Xueping, A.I.R. Herries, B. Kanning, P.S.C. Taçon, B. Zhende, D. Fink, Z. Yunsheng, J. Hellstrom, L. Yun, G. Cassis, S. Bing, S. Wroe, H. Shi, W.C. Parr, H.*Waddell (2016). Comparison of shape and morphological character based phylogenies of genus Homo* Page 10